\documentstyle[12pt,aasms4]{article}

\newcommand{\Vmag}{{\em V~}}

\newcommand{\Imag}{{\em I~}}
\newcommand{\Jmag}{{\em J~}}
\newcommand{\Hmag}{{\em H~}}
\newcommand{\Kmag}{{\em K~}}

\newcommand{\BVcol}{{\em B-V~}}

\newcommand{\VRcol}{{\em V-R~}}
\newcommand{\VKcol}{{\em V-K~}}
\newcommand{\JKcol}{{\em J-K~}}

\def\eg {{\rm e.g.}\ }
\def\ie {{\rm i.e.}\ }
\def\etal {{\rm et al.}\ }

\def\apj { ApJ}
\def\apjs {ApJS}
\def\apjl {ApJL}
\def\mnras {MNRAS}
\def\aj {AJ}

\def\pasp {PASP}

\def\littlesm{\ifmmode{\scriptscriptstyle m }
    \else{\hbox{$\scriptscriptstyle m $ }}\fi}
\def\littleprime{\ifmmode{\scriptscriptstyle \prime }
    \else{\hbox{$\scriptscriptstyle \prime$ }}\fi}
\def\littlecirc{\ifmmode{\scriptscriptstyle \circ }
    \else{\hbox{$\scriptscriptstyle \circ $ }}\fi}
\def\littless{\ifmmode{\scriptscriptstyle s }
    \else{\hbox{$\scriptscriptstyle s $ }}\fi}
\def\magm{\raise .9ex \hbox{\hskip-1pt\littlesm}}
\def\arcsec{\raise .9ex \hbox{\littleprime\hskip-3pt\littleprime}}
\def\arcmin{\raise .9ex \hbox{\littleprime}}
\def\degree{\raise .9ex \hbox{\littlecirc}}
\def\magpoint{\hbox to 1pt{}\rlap{\magm}.\hbox to 2pt{}}
\def\arcsecpoint{\hbox to 1pt{}\rlap{\arcsec}.\hbox to 2pt{}}
\def\arcminpoint{\hbox to 1pt{}\rlap{\arcmin}.\hbox to 2pt{}}
\def\degreepoint{\hbox to 1pt{}\rlap{\degree}.\hbox to 2pt{}}

\lefthead{Gieren, Fouqu\'e \& G\'omez}
\righthead{Cepheid PR and PL relations and LMC distance}

\begin{document}

\title{Cepheid Period-Radius and Period-Luminosity Relations
   and the Distance to the LMC}
\author{Wolfgang P. Gieren}
\affil{Universidad de Concepci\'on, Departamento de F\'{\i}sica, Casilla 4009,
   Concepci\'on, Chile; \\ email: wgieren@coma.cfm.udec.cl}

\author{Pascal Fouqu\'e}
\affil{Observatoire de Paris, Section de Meudon DESPA F-92195 Meudon CEDEX, 
   France; \\ European Southern Observatory, Casilla 19001, Santiago 19, Chile;
   \\ email: pfouque@eso.org}

\and 
\author{Mat\'{\i}as G\'omez}
\affil{P. Universidad Cat\'olica de Chile, Departamento de Astronom\'{\i}a y
   Astrof\'{\i}sica, \\ Casilla 104, Santiago 22, Chile; 
   \\ email: mgomez@astro.puc.cl}

\begin{abstract}

     We have used the infrared Barnes-Evans surface brightness technique to
derive the radii and distances of 34 Galactic Cepheid variables. Radius and 
distance results obtained from both versions of the technique are in excellent 
agreement. The radii of 28 variables are used to determine the period-radius 
relation. This relation is found to have a smaller dispersion than in previous 
studies, and is identical to the period-radius relation found by Laney \& 
Stobie from a completely independent method, a fact which provides persuasive 
evidence that the Cepheid period-radius relation is now determined at a very 
high confidence level. We use the accurate infrared distances to determine 
period-luminosity relations in the \Vmag, \Imag, \Jmag, \Hmag and \Kmag 
passbands from the Galactic sample of Cepheids. We derive improved slopes of 
these relations from updated LMC Cepheid samples and adopt these slopes to 
obtain accurate absolute calibrations of the PL relation. By comparing these 
relations to the ones defined by the LMC Cepheids, we derive strikingly 
consistent and precise values for the LMC distance modulus in each of the 
passbands which yield a mean value of $\mu_{\circ} (\rm LMC) = 18.46 \pm 0.02$.

     By analyzing the observed dispersions of the PL relations defined by the 
LMC and Galactic samples of Cepheids, we disentangle the contributions due to
uncertainties in the reddenings, in distance measurement and due to metallicity
effects, and we estimate the intrinsic dispersion of the PL relation with the 
Wesenheit function. Assuming that the Galactic Cepheid distances are typically 
accurate to $\pm 3\%$ (as shown in a previous Paper), and an intrinsic spread 
in [Fe/H] of $\sim 0.4$ dex among the Cepheids of our sample as obtained by 
Fry \& Carney, the observed dispersion of the Galactic Cepheid PL relation 
suggests a metallicity dependence of $\Delta \mu/\Delta {\rm [Fe/H]} \approx 
0.2$, about half the value suggested by Sasselov \etal from EROS data. When we 
apply this correction, the LMC distance modulus is increased to 
$18.52 \pm 0.06$ with most of this uncertainty being due to the adopted
metallicity correction.

     Our results show that the infrared Barnes-Evans technique is very 
insensitive to both Cepheid metallicity and adopted reddening, and therefore a 
very powerful tool to derive accurate distances to nearby galaxies by a 
direct application of the technique to their Cepheid variables, rather 
than by comparing PL relations of different galaxies, which introduces much 
more sensitivity to  metallicity and absorption corrections which are usually 
difficult to determine.

\end{abstract}


\section{Introduction}

     The determination of the radii and absolute magnitudes of Galactic Cepheid
variables, and the establishment of the corresponding period-radius and 
period-luminosity relations has occupied researchers for several decades. 
While the correct measurement of Cepheid radii, and hence a correct knowledge 
of the period-radius relation obeyed by Cepheid variables is important to 
determine the masses and other physical parameters of these variables, our 
ability to measure the distances to these stars critically determines our 
ability to scale the Universe out to several Megaparsecs and to lay the 
foundation to determine the Hubble constant. The period-radius relation may
also turn out to be very useful for the determination of pulsational 
parallaxes of Cepheids in galaxies whose distances are too large to allow to 
observe meaningful radial velocity curves of the variables, but still permit 
to obtain good light curves and periods.

     Significant progress in the determination of Cepheid distances and radii
has recently been made by Fouqu\'e \& Gieren \markcite{fou97} (1997, hereafter 
Paper~I) who have calibrated an infrared version of the Barnes-Evans (hereafter
BE) surface brightness technique using the \Kmag, \JKcol magnitude-color 
combination, as well as a version using the \Vmag, \VKcol combination which was
originally introduced by Welch \markcite{wel94} (1994), with an 
accurate zero point of the surface brightness-color relations determined from a
large set of interferometrically determined angular diameters of cool giants 
and supergiants which have become available over recent years. This new 
technique was applied to a sample of 16 Galactic open cluster Cepheids by 
Gieren, Fouqu\'e \& G\'omez \markcite{gie97} (1997, hereafter Paper~II), and it
was demonstrated in this paper that Cepheid radii and distances can be 
determined with a $\pm 3\%$ accuracy from both versions of the technique which 
yield identical results, within these small errors.

     In this paper, we extend our Cepheid sample to derive the period-radius
relation, and period-absolute magnitude relations in optical (\Vmag, \Imag) and
infrared (\Jmag, \Hmag, \Kmag) passbands from a statistically more significant 
number of Galactic Cepheid variables. As we will show in the forthcoming 
sections of this paper, our improved ability to measure the radii and distances
of individual Cepheids from our infrared technique, combined with a careful
selection of the stars adopted for this study, leads to relations of lower
dispersion than those obtained in any previous work.

     Since our Galactic Cepheid sample is not large enough to derive the slopes
of the PL relations in the different passbands with the highest possible 
accuracy, we adopt the approach to determine the slopes from the LMC Cepheids 
and use our Galactic Cepheid sample to set the zero points of the relations. 
These relations represent our best absolute calibrations of the Cepheid 
period-luminosity relations and will be derived in Section 5 of this paper. 
Comparing these Galactic PL relations to the corresponding relations defined by
the LMC Cepheids, we finally proceed to derive a new distance to the LMC. We 
also address the open question of the metallicity dependence of the PL relation
in this paper.

\section{Additional Radius and Distance Solutions}

     As in Paper~II for the cluster Cepheids, we adopted the infrared \Jmag, 
\Kmag photometry on the Carter system of Laney \& Stobie \markcite{lan92} 
(1992) for the additional Cepheids selected for this study. However, we have 
omitted a number of the Cepheids observed by these authors for our analysis, 
for one or several of the following reasons:

\begin{itemize}
\item [a)] insufficient number of \Jmag, \Kmag observations (N $< 20$)
\item [b)] insufficient number and/or quality of available radial velocity 
 observations
\item [c)] a red companion star present (which affects the observed infrared 
 light curves)
\item [d)] spectroscopic binary Cepheids with unknown orbital velocity curve 
 (\ie correction of the pulsational velocity curve for the orbital effect is 
 not possible)
\item [e)] Cepheids with extremely small light and velocity amplitudes (for 
 which our method produces unreliable radius and distance solutions)
\item [f)] double mode Cepheids
\end{itemize}

\placetable{tab1}
\begin{table}
\dummytable\label{tab1}
\end{table}

\placetable{tab2}
\begin{table}
\dummytable\label{tab2}
\end{table}

     These selection criteria left us with the additional 18 stars listed in 
Table~\ref{tab1}, and a total of 34 Cepheids adding the 16 Cepheids studied in 
Paper~II. While this sample is obviously smaller than the one studied by Laney 
\& Stobie \markcite{lan94} \markcite{lan95} (1994, 1995), it has the advantage 
of being a "clean" sample, freed from stars with expected systematic or 
enhanced random errors in their solutions, while at the same time being still 
large enough to provide good statistics in the relations we are going to 
investigate in this paper. As in Paper~II, we undertook a literature survey to 
determine the best available radial velocity and photometric \Vmag data for the
adopted Cepheids, where the database of D.~Welch \markcite{wel97} (1997) again 
provided valuable help. In Table~\ref{tab2}, the sources of the adopted data 
are listed. For all variables, we adopted the pulsation periods given by Laney 
\& Stobie \markcite{lan92} (1992). For the color excesses of all stars we 
adopted the mean values given in the Fernie \etal \markcite{fer97a} (1997) 
Galactic Cepheid database, which are listed with their uncertainties in 
Table~\ref{tab3}. We recall at this point that the BE technique is very 
insensitive to the values of the color excesses used in the analysis. For the 
conversion factor from radial to pulsational velocity we adopted the slightly 
period-dependent values resulting from the formula given by Gieren, Barnes 
\& Moffett \markcite{gie93a} (1993), as we did for the Cepheids analyzed in 
Paper~II.

\placetable{tab3}
\begin{table}
\dummytable\label{tab3}
\end{table}

     For each Cepheid variable we obtained two radius and distance solutions,
using the \Kmag, \JKcol and the \Vmag, \VKcol versions of the method, in 
exactly the same way as described in Paper~II. We recall that we adopt the 
inverse fits in all cases for the reasons stated and investigated in Paper~II. 
The resulting radii and distances are listed in Table~\ref{tab1}. For most 
Cepheids, the agreement between the two solutions is very good, but there are a
few exceptions. In order to decide on the final radius and distance to adopt 
for a given Cepheid, we used the plots of the linear and angular diameter 
variations vs. phase. We found that for the pure infrared \Kmag, \JKcol 
solutions, the agreement of the two curves is always excellent, while for the 
\Vmag, \VKcol solutions the agreement is not so good for a fraction of the 
Cepheids, specially for those with the longest pulsation periods in the sample.
We have seen a similar effect in the Cepheids studied in Paper~II and 
identified as the most likely cause a slight phase mismatch between the \Vmag 
and \Kmag light curves (which were not obtained simultaneously) used in the 
\Vmag, \VKcol analyses due to an increasing tendency of period variability in 
the long-period Cepheids, a problem which does not exist in the \Kmag, \JKcol 
solutions because here all photometric data were obtained contemporaneously. 
There is also a possibility that the \Vmag, \VKcol infrared BE technique begins
to work less well for the very extended, luminous Cepheids while it still works
well at pure infrared wavelengths for these stars. For those Cepheids where 
problems of this kind are clearly visible in their \Vmag, \VKcol solutions, we 
adopted as the final radius and distance the one coming from the \Kmag, \JKcol 
solution alone. For the other (and the majority of) Cepheids for which this 
problem does not exist, the final adopted radii and distances are the weighted 
means of both solutions. We give the final, adopted radii and distances for all
Cepheids, including those studied in Paper~II, in Table~\ref{tab4}. Note that 
the uncertainties in these adopted radii and distances are calculated according
to the formulae given in the Appendix of Quintana \etal \markcite{qui94} 
(1994), \ie they take into account not only the individual uncertainty of each 
solution, but also the difference between the two solutions, which may evidence
systematic errors. We remark that, had we chosen to adopt the mean values from 
both solutions for all Cepheids (except the three longest-period stars of the 
sample which will be discussed later), the conclusions of this paper would not 
change in any significant way.

\placetable{tab4}
\begin{table}
\dummytable\label{tab4}
\end{table}

     In Fig.~\ref{fig1}, we compare the radii determined from the two different
infrared BE techniques. All stars having both solutions (32) are included in
this Figure. To calculate the mean ratio $R_{\rm V-K}/R_{\rm J-K}$, we exclude 
the three shortest-period stars (unreliable \Kmag, \JKcol solution) and 
GY~Sge, for which the \Vmag, \VKcol solution is obviously unsuccessful. We then
find the mean ratio $R_{\rm V-K}/R_{\rm J-K}$ to be $0.998 \pm 0.013$ (27 
stars). If we compare the radii only for those Cepheids of the sample for which
we adopted the mean of both solutions, the corresponding value is again 
$1.00 \pm 0.01$. This very clearly demonstrates that both infrared BE 
techniques produce identical radius results, within a very small error and 
independent of pulsation period. We note, however, that there is a tendency in 
Fig.~\ref{fig1} for the scatter to increase toward longer periods, which we 
attribute to the problem of the non-simultaneous \Vmag, \Kmag photometry in the
\Vmag, \VKcol solutions, as discussed above.

\placefigure{fig1}

     In Fig.~\ref{fig2}, we show the same comparison for the distances. This 
time, the corresponding ratios are $0.983 \pm 0.013$ and $0.98 \pm 0.01$, 
respectively, and as in the case of the radii, there is clearly no evidence for
this ratio to vary systematically with period. In Paper~II we had found a value
of 0.97, so the inclusion of more Cepheids in the comparison has brought the 
ratio even closer to unity. While one might speculate that the very small 
offset between the distances derived from the two methods might be real, it is 
clearly within the possible systematic errors of either method discussed in 
Paper~II, thus justifying as an optimum choice to average the distances, as 
well as the radii, from both infrared BE techniques, as long as there are no 
clear reasons in particular cases to exclude one of the solutions.

\placefigure{fig2}

\section{The Period-Radius Relation}

     When we plot in Fig.~\ref{fig3} the period-radius relation from the radius
data in Table~\ref{tab4}, we note that the three longest-period Cepheids of our
sample (SV~Vul, GY~Sge and S~Vul) lie clearly above the very tight relation 
defined by all the other stars. While we cannot exclude the possibility that 
the radii of these stars are correct, we prefer to conclude that our radius 
determinations overestimate the true radii in these cases and eliminate these 
stars in establishing the period-radius relation. There are several 
justifications for suspecting a problem with these very long-period stars: 
first, the pronounced disagreement between the \Kmag, \JKcol and the \Vmag, 
\VKcol solutions for GY~Sge, which might imply that part of the problem is also
with the 
(adopted) \Kmag, \JKcol solution, and not only with the \Vmag, \VKcol solution,
for very luminous supergiants with very extended atmospheres; second, there is
clearly a problem of a variable period for SV~Vul (Bersier \etal 
\markcite{ber94} 1994; Berdnikov \markcite{ber97b} 1997), and GY~Sge and S~Vul 
(Berdnikov \markcite{ber97b} 1997) which may have caused a systematic error in 
our solutions. Particularly in the case of GY~Sge and S~Vul, we were not able 
to find a single period which represents satisfactorily the different sets of 
photometric data available in the literature. Finally, the range
of periods for which our infrared BE techniques are calibrated (see Paper~I) is
up to 40 days, and while there is no particular reason to suspect that the 
calibration is different for Cepheids of longer periods, it is just these three
stars which lie outside of the calibration range. There is clearly less 
confidence in the radius and distance solutions of these Cepheids, and we 
therefore feel that it is wise to exclude them from the calibration of the PR 
relation. We therefore do not think that the curvature in the PR relation
(Fig.~\ref{fig3}) which may be suggested by these stars is real.

     On the other extreme of the period spectrum, there is a possible ambiguity
as to the pulsation modes of the three shortest-period stars in our sample, 
which are EV~Sct, SZ~Tau and QZ~Nor. Evidence for all of these Cepheids to be 
first overtone pulsators has been brought forward repeatedly in the literature,
and is supported by the near-sinusoidal shapes of their light curves, but an 
uncertainty with regard to the mode identification remains, and we therefore 
choose to omit these stars from the discussion of the PR relation as well.

     From the remaining 28 Cepheids of our sample, we find

\begin{equation}
 \log R = 0.750 \  (\pm 0.024) \  \log P + 1.075 \  (\pm 0.007),
\end{equation}

as the resulting period-radius relation, with a dispersion of $\sigma = 0.036$
and a correlation coefficient of $\rho = 0.987$. This relation is plotted in 
Fig.~\ref{fig3}. It is identical to the one found by Laney \& Stobie 
\markcite{lan95} (1995, hereafter LS95) from an application of the maximum 
likelihood method to a somewhat larger Cepheid sample which includes the 
present one, which is

\begin{equation}
 \log R = 0.751 \  (\pm 0.026) \  \log P + 1.070 \  (\pm 0.008),
\end{equation}

\noindent
with a dispersion of $\sigma = 0.051$ (we note that LS95 have used a constant 
p-factor of 1.36 which corresponds to our value for intermediate-period stars. 
The same choice would have made the slope of our PR relation steeper, but by a 
very small amount well within the errors). While many of the data of the 
individual Cepheids, in particular the infrared photometry, are shared by both 
studies, the methods to derive the radii are completely independent, and we 
thus feel that the perfect agreement of our result with LS95 provides extremely
persuasive evidence that we have now established the true period-radius 
relation obeyed by classical Cepheids in our Galaxy, with a very high degree of
confidence and within the small errors of the coefficients stated above. We 
also note that in a direct star-to-star comparison of our radii to those 
derived by LS95 (both radii normalized to the same p-factor), the radii agree 
to better than 10 percent in all cases except one (CV~Mon), and in most cases 
to better than 5 percent, without any dependence on period, which reassures 
that we are now able to measure very accurate radii of individual Cepheid 
variables using infrared photometry and both, the Barnes-Evans, and the maximum
likelihood technique.

\placefigure{fig3}

    Two other recent efforts to calibrate the Cepheid PR relation are the work
of Ripepi \etal \markcite{rip97} (1997), and of Krockenberger \etal 
\markcite{kro96} (1996). While Ripepi \etal use a modified version of the CORS 
method (Caccin \etal \markcite{cac81} 1981) which makes use of two optical 
color indices (\BVcol and \VRcol), Krockenberger \etal have devised a method of
the Baade-Wesselink type which makes use of the Fourier coefficients of the 
observables and have applied it using again optical photometry (on the Geneva 
system) in their analysis. In both studies, the resulting slopes of the PR 
relation are much shallower than our result, close to 0.60. We suspect that in 
both studies the problem is not with the techniques, but with the use of 
optical photometry in the application to Cepheid variables which is not able to
provide correct estimates of the surface brightness, apparently even in the 
case of the CORS method which tries to remedy the problems by the introduction 
of a second color index. On the other hand, it is interesting to note that the 
period-radius relation derived by Gieren, Barnes \& Moffett \markcite{gie89b} 
(1989) which is based on optical (\VRcol) Barnes-Evans radii of 100 galactic 
variables is very close to the relation found from infrared photometry. LS95 
have tried to explain this as a coincidence in the selection of the variables, 
but this seems unlikely in view of the large sample used by Gieren \etal. The 
results in Paper~II in which we have also derived the radii from our newly 
calibrated optical version of the BE method shed some light on this question. 
They indicate that one can have large systematic errors in individual optical 
BE radii (up to $\sim 30$ percent), but since these systematic errors can 
apparently have either sign it is possible that they cancel out, to a large 
degree, in the determination of a mean period-radius relation based on a large 
number of stars. This possibly explains why the Gieren \etal \markcite{gie89b} 
(1989) PR relation is close to the true relation found from infrared 
photometry, but exhibits a much larger dispersion than the relation found in 
this paper, due to a strong contribution of observational scatter to the total 
dispersion.

     The dispersion of our infrared-based Cepheid PR relation is smaller than 
in any previous study and is close to the dispersion expected from the finite 
width of the Cepheid instability strip, with the mass being the third parameter
in the full period-radius-mass relationship. Given that the mass depends 
sensitively on the radius, approximately as $M \sim R^{\, 2.5}$, and that 
individual radii of Cepheids could not be determined with an accuracy better 
than $\sim 10$ percent, it has hitherto not been possible to derive good
individual masses for Cepheid variables via the PRM relation (\eg Gieren
\markcite{gie89a} 1989). However, the Cepheid radii based on the new infrared 
BE technique are on average accurate to 3 percent and should therefore allow, 
for the first time, to derive Cepheid masses from the pulsational PRM relation 
being accurate to better than 10 percent. This will result in important 
progress in the determination of this most fundamental stellar parameter for 
supergiant stars.

\section{Optical and Near-Infrared Period-Luminosity Relations}

     We now turn to the discussion of the Cepheid period-luminosity relation in
the optical \Vmag, \Imag and in the near-infrared \Jmag, \Hmag and \Kmag 
passbands, which we will derive from our new infrared distances to our Galactic
Cepheid calibrating sample. In order to convert the distances into absolute 
magnitudes, we have to adopt mean magnitudes and absorption corrections for the
variables. For all variables in our sample, we have adopted the intensity mean 
magnitudes in \Vmag, \Jmag, \Hmag and \Kmag as given by Laney \& Stobie 
\markcite{lan93} (1993). The intensity mean magnitudes in the \Imag band were 
derived from Caldwell \& Coulson \markcite{cal87} (1987), adding a constant 
$-0.03$ mag correction to their magnitude means to convert them into intensity 
means, a correction which was found appropriate from tests on several Cepheids 
of different periods and light amplitudes. All these data, together with the 
adopted color excesses of the stars from the Fernie \etal database and the 
intensity mean \BVcol colors (again from Laney \& Stobie \markcite{lan93} 1993)
are given in Table~\ref{tab3}.

     The absorption corrections were calculated from 
$A_{\lambda} = R_{\lambda} \  E(B-V)$ using the following expressions:

\begin{eqnarray}
R_{\rm V} & = & 3.07 + 0.28 \  (B-V)_{\circ} + 0.04 \  E(B-V) \\
R_{\rm I} & = & 1.82 + 0.205 \  (B-V)_{\circ} + 0.022 \  E(B-V) \\
R_{\rm J} & = & 0.764 \\
R_{\rm H} & = & 0.450 \\
R_{\rm K} & = & 0.279 
\end{eqnarray}

The (constant) $R$ values for the infrared passbands were adopted from the 
work of Laney \& Stobie \markcite{lan93} (1993), while the expressions for the 
ratios of total to selective absorption in the optical \Vmag and \Imag bands 
were adopted from Caldwell \& Coulson \markcite{cal87} (1987), and Laney \& 
Stobie \markcite{lan93} (1993). We then calculated the absolute magnitudes in 
the different passbands from the absorption-corrected intensity mean magnitudes
and the true distance moduli of the stars as determined from their distances 
given in Table~\ref{tab4}. The resulting absolute magnitudes in the optical 
\Vmag and \Imag bands, together with their uncertainties due to the combined 
effect of distance and reddening uncertainty, are given in Table~\ref{tab5}, 
while in Table~\ref{tab6} we give the infrared absolute magnitudes of the stars
and their uncertainties. From these data, we derived period-absolute magnitude 
relations by least-squares fits to the same subset of 28 Cepheids discussed in 
the previous Section, with the results for slope, zero point and dispersion of 
the relations as given in Table~\ref{tab7}. We display these relations in 
Figs.~\ref{fig4} to \ref{fig8}. The observed dispersion in the PL relations 
defined by our Galactic Cepheid sample decreases from 0.21 mag in the \Vmag 
band to 0.17 mag in \Kmag, and the corresponding observed total widths of the 
relations decrease from about 0.7 mag in \Vmag to 0.5 mag in \Kmag.

\placetable{tab5}
\begin{table}
\dummytable\label{tab5}
\end{table}

\placetable{tab6}
\begin{table}
\dummytable\label{tab6}
\end{table}

\placetable{tab7}
\begin{table}
\dummytable\label{tab7}
\end{table}

\placefigure{fig4}
\placefigure{fig5}
\placefigure{fig6}
\placefigure{fig7}
\placefigure{fig8}

     In order to judge the improvement in the accuracy of our measurement of
Cepheid distances, it is instructive to compare the present Galactic \Vmag band
PL relation to the one derived from the visual surface brightness technique by 
Gieren \etal \markcite{gie93a} (1993). The total width, observed to be 
$\sim 1.1$ mag in the Gieren \etal \markcite{gie93a} (1993) relation, has now 
decreased to $\sim 0.7$ mag, and the dispersion of the relation has decreased 
from 0.3 to 0.2 mag. This huge improvement is almost exclusively due to the 
improvement in distance measurement since there has been little change in the 
adopted absorption corrections.

     One very important point which could not yet be properly addressed in 
Paper~II refers to the metallicity sensitivity of our infrared method to 
measure Cepheid distances. From model atmosphere studies, like the one of 
Bell \& Gustafsson \markcite{bel89} (1989), we can conclude that the effect of 
a changing metallicity on the infrared colors of Cepheid variables is very 
small, and thus we expect that the infrared BE technique yields radius and 
distance results which are very little dependent on the metallicities of the 
stars. However, it is good to check this prediction empirically, and we can do 
so. One way to do this is to look at the dispersion of the PL relation. First, 
it is important to note that our Cepheid sample is by no means homogeneous with
respect to metallicity. The excellent study of Fry \& Carney \markcite{fry97} 
(1997) of the chemical abundances of a large fraction of Galactic open cluster 
Cepheids, many of which are in our present sample as well, has shown that there
is clearly a genuine metallicity spread among these stars, which amounts to 
$\sim 0.4$ dex in [Fe/H] and is roughly bracketing the solar metallicity. While
the amount of this spread may be surprisingly large, there has been evidence 
before (\eg Giridhar \markcite{gir86} 1986) that Cepheids appear to become, in 
a systematic way, more metallicity-poor with increasing distance from the 
Galactic center. However, as shown by Fry \& Carney, their very accurate and 
homogeneous metallicity determinations of the open cluster Cepheids do not 
support Giridhar's earlier conclusion; the plot of metallicity versus 
galactocentric distance turns out to be basically a scatter plot. Thus we do 
not expect that the galactocentric distances of our Cepheids (as a supposed 
crude indicator of metallicity) correlate with the residuals of the absolute 
magnitudes from the mean \Kmag band PL relation (which are the best suited for 
the problem because they are almost unbiased by possible errors in the color 
excesses), and effectively this is not the case. It is therefore not possible 
to detect a metallicity dependence of our distance measuring method in this 
way. However, the small dispersion of the PL relation itself already tells us 
that, given the relatively large $\sim 0.4$ dex spread among the metallicities 
of the Cepheids of our sample, the distances cannot depend strongly on 
metallicity, because this would have to show up as an additional, significant 
observational scatter in the PL relation. While there probably is a 
contribution to the dispersion which is metallicity-related (see discussion in 
Section~5), this small effect is almost certainly due to a (slight) metallicity
dependence of the Cepheid absolute magnitudes, and not an artefact of the 
technique we use to measure the distances. The small size of any 
metallicity-related effect is demonstrated by Fig.~\ref{fig9} where we have 
plotted the $M_{\rm K}$ residuals against the [Fe/H] values for the eight stars
in common with the Fry \& Carney open cluster Cepheid sample -- clearly, there 
is no detectable correlation of these residuals with metallicity.

\placefigure{fig9}

\section{Improved Absolute Calibration of Cepheid PL Relations and the Distance
   to the LMC}

     Since our Galactic Cepheid sample is relatively small, we can obtain a 
more accurate determination of the slopes of the PL relations in the different 
passbands by using Magellanic Cloud Cepheids. This assumes, of course, that the
slope of the PL relation (in all passbands) is universal, a question which 
still awaits an exhaustive empirical check, but which has some supportive 
evidence to the moment (\eg Musella \etal \markcite{mus97} 1997; 
Sasselov \etal \markcite{sas97} 1997) and seems to be firmly supported by 
theoretical expectations (\eg Stothers \markcite{sto88} 1988). While the LMC 
has a relatively small intrinsic depth in the line of sight and there is the 
possibility to correct LMC Cepheid magnitudes for the tilt of the LMC bar 
(Caldwell \& Laney \markcite{cal91} 1991), the situation is more complicated 
for the SMC Cepheids, and for this reason we will derive PL relations only from
the Cepheids in the Large Magellanic Cloud. We will then fit the slopes found 
from the LMC Cepheid samples to the Galactic relations and this way obtain our 
best absolute calibrations of the Cepheid PL relations in the \Vmag, \Imag, 
\Jmag, \Hmag and \Kmag passbands. Comparison of these relations to the 
corresponding relations in the LMC will yield a LMC distance value in each of 
these passbands, from which a best mean LMC distance will be derived. While 
this distance still bears some dependence on metallicity and the adopted LMC 
Cepheid reddening corrections, we will present evidence for this having only a 
small effect (in the order of a few hundredths of a mag) on our adopted LMC 
distance.

\subsection{LMC Cepheid period-luminosity relations in \Vmag, \Imag, \Jmag, 
   \Hmag and \Kmag}

     A first and very important step is to define the LMC Cepheid samples to 
adopt for our purpose. This task was greatly facilitated by the LMC Cepheid 
data base put at our disposal by J.A.R. Caldwell. After inspecting the 
available data and their quality, we decided, in the \Vmag and \Imag bands, to 
use Tanvir's \markcite{tan97} (1997) sample of 53 LMC Cepheids with photometry 
in both bands and $\log P < 1.8$. We have improved the periods and intensity 
means in \Vmag and \Imag for 9 Cepheids of this sample from new, high-quality 
light curves obtained for these variables by Moffett \etal \markcite{mof97} 
(1997). We have also used individual reddenings for 25 Cepheids from the 
Caldwell data base which were obtained as described by Laney \& Stobie 
\markcite{lan94} (1994). For the remaining Cepheids, we have adopted Caldwell's
average reddening value of $E(B-V) = 0.07$, except for the Cepheids in the 
field of the cluster NGC~1850 for which we adopted 0.15, as recommended by Sebo
\& Wood \markcite{seb95} (1995). Correction for absorption has been made 
according to the following equations, where the mean $R_{\rm V}$ value 
appropriate to Cepheid colors was taken from Gieren \& Fouqu\'e 
\markcite{gie93b} (1993) and the value $A_{\rm I}/A_{\rm V} = 0.592$ from 
Tanvir \markcite{tan97} (1997):

\begin{eqnarray}
 \langle V_{\circ} \rangle = \langle V \rangle - 3.26 \  E(B-V) \\
 \langle I_{\circ} \rangle = \langle I \rangle - 1.93 \  E(B-V)
\end{eqnarray}

The adoption of individual reddenings slightly improves the dispersion about 
the mean PL relations, from 0.233 to 0.204 in \Vmag, and from 0.164 to 0.150 in
\Imag. Determination of accurate individual reddenings to each Cepheid in the 
sample should allow to recover the smaller dispersion of the 
reddening-independent Wesenheit function (see below). Small corrections for the
tilt of the LMC against the plane of the sky have not been applied because they
did not significantly improve the dispersion. For the sake of clarity, we list 
in Table~\ref{tab8} the final sample for the LMC Cepheid PL solutions in the 
\Vmag and \Imag bands, with the adopted values of the periods, 
absorption-corrected intensity mean magnitudes and extinctions. The PL 
relations resulting from least-squares fits to these data are given in 
Table~\ref{tab10}. The slopes we find are undistinguishable from Tanvir's 
\markcite{tan97} (1997) values.

\placetable{tab8}
\begin{table}
\dummytable\label{tab8}
\end{table}

     Unfortunately, the \Jmag, \Hmag and \Kmag band coverage of Tanvir's sample
of Cepheids is far from complete, and we therefore prefer using the Laney \& 
Stobie \markcite{lan94} (1994) sample as a starting point for the infrared 
bands. This sample contains 19 LMC Cepheids with good infrared light curves and
33 with few-phase IR data, adopted from Welch \etal \markcite{wel87} (1987) and
transformed to the Carter system. In fact, 7 other Cepheids have IR data from 
Welch \etal, but have not been retained by Laney \& Stobie. As they do not seem
to increase the dispersions of the infrared PL relations, we prefer to adopt 
the complete sample of 59 Cepheids, and transform the Welch \etal data 
following the Laney \& Stobie precepts for system conversion and dereddening. 
Extinction values for these 7 additional stars have also been provided by 
Caldwell. Again, no tilt corrections were applied, as they did not improve the 
dispersions about the PL relations. We list in Table~\ref{tab9} the final 
sample of LMC Cepheids adopted for the infrared PL solutions with the adopted 
periods, absorption-corrected intensity mean magnitudes and extinctions. The PL
relations in \Jmag, \Hmag and \Kmag were derived from least-squares fits to 
these data and are given in Table~\ref{tab10}. They significantly differ from 
the PL relations given in Laney \& Stobie \markcite{lan94} (1994), because 
these authors mixed LMC, SMC and Galactic Cepheid samples in their solutions. 
Note that the dispersion of the \Kmag band PL relation is about the same as for
the \Vmag - \Imag Wesenheit function (see below), which is not surprising since
both relations are almost unaffected by reddening corrections.

\placetable{tab9}
\begin{table}
\dummytable\label{tab9}
\end{table}

\placetable{tab10}
\begin{table}
\dummytable\label{tab10}
\end{table}

\subsection{The \Vmag -- \Imag Wesenheit function and the observed PL 
   dispersions}

     In order to find out the intrinsic dispersion of the PL relation, and the 
contribution of uncertainties in reddening and distance to the observed, total 
dispersions in the different passbands, we constructed the \Vmag -- \Imag 
Wesenheit function for the LMC Cepheid sample. In this sample, the contribution
of distance errors to the observed dispersion is negligible (since all stars 
are basically at the same distance), but there will be a contribution due to 
errors in the adopted absorption corrections. These errors are removed to a 
large extent if one uses the reddening-independent Wesenheit function defined 
as

\begin{equation}
 W = V_{\circ} - R \  (\langle V_{\circ} \rangle - \langle I_{\circ} \rangle)
   = V - R \  (\langle V \rangle - \langle I \rangle)
\end{equation}

where $R$ is defined as $A_{\rm V}/(A_{\rm V} - A_{\rm I})$ and is obtained as 
the slope of the fit of the \Vmag band PL relation residuals to the residuals 
from a mean $\langle V_{\circ} \rangle - \langle I_{\circ} \rangle$ 
period-color relation. Using the data of Table~\ref{tab8}, we determined 
$R = 2.34 \pm 0.22$, which is close to the expected value of 2.45 which 
corresponds to reddening correction. Using this value in the Wesenheit function
and plotting $W$ against $\log P$, we find a relation whose dispersion has 
decreased to 0.113 mag, which we might then interpret as the intrinsic 
dispersion of the \Vmag band PL relation. A very similar Wesenheit relation is 
displayed as Fig.~3 in Tanvir \markcite{tan97} (1997). A corresponding plot of 
the \Kmag band Wesenheit function against $\log P$ yields a relation whose rms 
dispersion is 0.114 mag, and there is no gain as compared to the $K_{\circ}$ PL
relation, which is expected since at \Kmag the corrections for absorption for 
our LMC Cepheid sample are negligible. This finding is consistent with our 
interpretation of the intrinsic dispersion of the Cepheid PL relation being 
$\sim 0.11$ mag, and this value seems to remain much the same as going from 
\Vmag to \Kmag.

     Unlike the LMC Cepheid sample, the dispersions observed in the Galactic
Cepheid sample PL relations do contain a significant contribution due to errors
in the distance measurements of the individual Galactic Cepheids. Building a 
\Vmag -- \Imag Wesenheit function for the Galactic Cepheid sample in the same 
way as done above for the LMC sample, the $W$ versus $\log P$ relation is found
to have a dispersion of 0.17, as compared to the 0.21 mag dispersion shown by 
the Galactic \Vmag band PL relation, and very similar to the dispersion found 
in the \Kmag band PL relation where contributions from reddening errors are 
also negligible. Since we know the intrinsic dispersion of the \Vmag band PL 
relation from the LMC sample (see above), the remaining dispersion we observe 
should be due to errors in the distances, and perhaps to a metallicity-related 
effect. This remaining dispersion is 0.12 mag, and corresponds to an error of 
$\pm 5\%$ in the distances if it is completely due to distance uncertainties. 
From the results of Paper~II we know, on the other hand, that the expected 
uncertainty of a typical distance is $\sim \pm 3\%$, so there might be a small,
metallicity-related contribution in the same order which is probably not due to
a dependence of our technique on metallicity, but rather to a slight systematic
dependence of Cepheid absolute magnitudes on metallicity as found by the EROS 
results (Sasselov \etal \markcite{sas97} 1997) which seems to amount to 
$\sim 0.06$ mag in the metallicity range of $\Delta {\rm [Fe/H]} \approx 0.4$ 
dex covered by the Cepheids of our Galactic sample, and which is small enough 
to be hidden in Fig.~\ref{fig9} (and smaller than the metallicity dependence of
the \Vmag band PL relation suggested by Sasselov \etal \markcite{sas97} 1997). 
As a note of caution, this (rough) estimate of the metallicity dependence of 
the PL relation assumes that the intrinsic width of the instability strip is 
the same in LMC and the Galaxy.

\subsection{The distance of the LMC and the absolute calibration of the PL 
   relation in \Vmag, \Imag, \Jmag, \Hmag and \Kmag}

     The determination of an accurate and reliable distance to the LMC is a 
fundamental step in the extragalactic distance scale. Recent results from 
various distance indicators show that the range of values for $\mu_{\circ}$ 
(LMC) is from about 18.3 from Hipparcos proper motion-based RR Lyrae distances 
(Fernley \etal \markcite{fer97b} 1997) to 18.70 based on Hipparcos 
trigonometric parallax measurements of a sample of nearby Galactic Cepheids 
(Feast \& Catchpole \markcite{fea97} 1997), with a SN~1987A ring upper limit on
the LMC distance modulus of 18.44 lying between these extremes (Gould \& Uza 
\markcite{gou97} 1997). As a consequence, the true LMC distance is still 
uncertain at the 20\% level which is a very unsatisfactory situation. 
Furthermore, Cepheid-based LMC distance moduli tied to the ZAMS-fitting method 
and a traditional Pleiades distance modulus of 5.57 have now to be revised in 
accordance with the new Pleiades distance modulus value of 5.33 obtained from 
Hipparcos data (Mermilliod \etal \markcite{mer97} 1997; van Leeuwen 
\& Hansen~Ruiz \markcite{vle97} 1997) which brings these LMC distance estimates
(\eg Laney \& Stobie \markcite{lan94} 1994) close to 18.3, a value similar to 
the one derived from the RR Lyraes. To make things even worse, the Hipparcos 
results on several nearby open cluster distances have cast serious doubts on 
the small intrinsic dispersion among the locations of open cluster main 
sequences on which this method rests (Mermilliod \etal \markcite{mer97} 1997). In view of this situation a determination of the distance to the LMC from yet 
another independent method like the one used by us is clearly very important.

     Although the slopes of the LMC PL relations in Table~\ref{tab10} differ 
from the best fit slopes from our 28 Galactic calibrators, we attribute this 
difference to small number statistics. Indeed, looking at Figs.~\ref{fig4} to 
\ref{fig8} where the galactic data are displayed with the LMC relations 
superimposed, we see that the difference in the slopes may not be significant. 
As the LMC samples are larger and the dispersion of the LMC relations are 
smaller than that of their Galactic counterparts, we force the LMC slopes to 
the Galactic sample to establish an absolute zero point of the PL relations, 
and an absolute distance to the LMC in each band.

     In order to take account of the variable accuracies of the distances of 
our Galactic calibrating Cepheids, we have taken a weighted mean of the LMC 
distance moduli calculated from each Galactic Cepheid, in each band. The 
uncertainty of this weighted mean is the quadratic sum of the weighted 
dispersion divided by the square root of the number of Cepheids (28 in all 
bands except \Imag, with 27), and of the mean error of the intercept of the 
corresponding LMC PL relation. Results are given in the last column of 
Table~\ref{tab10} for each band, and the agreement among the distance moduli 
derived from the different bands is striking. From a weighted mean of these 
values, we obtain as the final distance modulus of the LMC

\begin{displaymath}
 \langle \mu_{\circ} {\rm (LMC)} \rangle = 18.46 \pm 0.02
\end{displaymath}

Subtracting this value from the intercept of the LMC PL relation in each band 
yields our adopted absolute calibrations, which now do not depend on any 
assumed LMC distance and mean extinction, and may be used to calibrate, for 
instance, results from HST in external galaxies if the metallicity is not too 
far from solar or LMC values. The uncertainty of the absolute intercept of our 
adopted PL relations is the quadratic sum of the corresponding uncertainty in 
the LMC PL relation intercept, and of the mean error of the LMC distance 
modulus. With this, our final absolute calibrations of the Cepheid PL 
relations in the various bands are then:

\begin{eqnarray}
 M_{\rm V} & = & -2.769 \  (\pm 0.073) \times (\log P - 1.0) - 4.063 \  (\pm 0.034),
 \label{eq:defMv} \\
 M_{\rm I} & = & -3.041 \  (\pm 0.054) \times (\log P - 1.0) - 4.767 \  (\pm 0.029),
 \label{eq:defMi} \\
 M_{\rm J} & = & -3.129 \  (\pm 0.052) \times (\log P - 1.0) - 5.240 \  (\pm 0.028),
 \label{eq:defMj} \\
 M_{\rm H} & = & -3.249 \  (\pm 0.044) \times (\log P - 1.0) - 5.628 \  (\pm 0.026),
 \label{eq:defMh} \\
 M_{\rm K} & = & -3.267 \  (\pm 0.042) \times (\log P - 1.0) - 5.701 \  (\pm 0.025).
 \label{eq:defMk}
\end{eqnarray}

     Comparison of these relations to the multiwavelength PL solutions of 
Madore \& Freedman \markcite{mad91} (1991) shows that the slopes of the \Vmag 
and \Imag band relations are almost identical, but that the present zero points
are 0.10 mag fainter in both bands. The dispersions of the present relations 
are significantly smaller, and so are the uncertainties on the coefficients of 
Eqs.~\ref{eq:defMv} and \ref{eq:defMi}. The \Jmag, \Hmag and \Kmag band 
relations of Madore \& Freedman yield almost identical absolute magnitudes at 
$\log P = 1.0$, but their slopes are significantly larger than ours, as are the
dispersions of their relations. We attribute this difference to the larger 
number of stars and improved photometry we have been able to use in our 
solutions.

     As noted before, our way of deriving the LMC distance modulus assumes that
there is no metallicity effect on the PL relation. To correct for the mean 
metallicity difference between the Galactic and the LMC Cepheid samples of 
$\sim 0.3$ dex, we might adopt the small 0.02 mag shift found by Laney \& 
Stobie \markcite{lan94} (1994), or the larger 0.14 mag shift following from the
results of Sasselov \etal \markcite{sas97} (1997). On the other hand, since the
metallicity spread in our Galactic sample which is close to the systematic 
metallicity difference between the Galaxy and LMC seems to introduce a 
$\sim 0.06$ mag shift in the absolute magnitudes, we prefer to adopt this value
to allow for the metallicity dependence of our LMC distance modulus. Doing so, 
we obtain as a metallicity-corrected distance modulus of the LMC

\begin{displaymath}
 \mu_{\circ} {\rm (LMC)} = 18.52 \pm 0.06
\end{displaymath}

where a $\pm 0.06$ mag uncertainty on the metallicity-induced shift has been 
assumed which determines almost completely the total uncertainty of the LMC 
distance modulus. Obviously, one would like to reduce the uncertainty of the 
metallicity correction to the LMC distance modulus, and a very promising way of
doing this is to apply the infrared Barnes-Evans method directly to LMC 
Cepheids, taking advantage of the fact that distances measured with this method
are almost completely independent of both absorption and metallicities of 
the target Cepheids. Such a program is currently underway and should yield the
true distance modulus of the LMC with an accuracy of $\sim 0.02$ mag.

\section{Conclusions}

     We have determined the radii and distances of 34 Galactic Cepheid 
variables from the infrared Barnes-Evans surface brightness technique of 
Fouqu\'e \& Gieren \markcite{fou97} (1997). We find that the two versions of 
the technique produce radii which agree to better than 1\% and distances which 
agree at the 2\% level, which is within the total uncertainty of both versions 
of the method. The radius data are used to construct a period-radius relation 
which shows a dispersion of only $\pm 0.036$ in $\log R$ about the mean 
relation, smaller than in any previous determination. We use the infrared 
distances of the variables to determine the period-luminosity relations in the 
optical \Vmag and \Imag, and in the near-infrared \Jmag, \Hmag and \Kmag 
passbands, and again find smaller dispersions in any of these relations than in
previous studies. In order to obtain absolute calibrations of the PL relation 
in each of the passbands which are as accurate as possible, we determine the 
slopes from larger LMC Cepheid samples which show PL relations of smaller 
dispersions, due to a negligible contribution of distance uncertainties to the 
observed dispersions. Adopting the slopes defined by the LMC samples, we then 
use the Galactic Cepheid sample to determine the absolute zero point of the PL 
relation in each passband. Comparing the Galactic PL relations to the ones 
defined by the LMC samples, we find values of the true, absorption-corrected 
distance modulus of the LMC in each band which show remarkable agreement among 
themselves and are very close to the "classical" value of the LMC distance 
modulus of 18.50 adopted by the HST Distance Scale Key Project team (\eg 
Madore \& Freedman \markcite{mad97} 1997).

     We use Wesenheit functions to disentangle the effects of reddening and 
distance uncertainty, and of the intrinsic dispersion (due to the finite 
width of the Cepheid instability strip) on the total observed 
dispersions in the LMC and Galactic Cepheid samples. From this, we conclude 
that the intrinsic dispersion of the PL relation is 0.11 mag, without a 
significant variation between the \Vmag and \Kmag bands, and we are able to 
estimate the contribution of the combined effect of distance uncertainty 
and a possible systematic effect of metallicity on the observed dispersion of 
the Galactic sample. From this, and from the knowledge that the [Fe/H] spead of
our sample is $\sim 0.4$ dex, we estimate that the metallicity effect on the 
distance modulus is $\Delta \mu / \Delta {\rm [Fe/H]} \approx 0.2$, about half 
the value suggested by the analysis of Sasselov \etal based on the EROS data. 
We note that this estimate relies on the excellent spectroscopic metallicity 
determinations of Galactic Cepheids of Fry \& Carney \markcite{fry97} (1997), 
and on the $\pm 3\%$ accuracy figure in distance measurement with the infrared 
Barnes-Evans technique found in Paper~II. A larger uncertainty in the 
distance measurement would imply a lower metallicity sensitivity of 
Cepheid absolute magnitudes. The infrared surface brightness technique itself 
appears to yield distances which are almost metallicity-independent, in 
agreement with model atmosphere predictions; this feature, together with the 
insensitivity of the method to adopted absorption corrections, makes it an 
almost ideal instrument to determine the true distances to several nearby 
galaxies with a high accuracy, and thus make a very important contribution 
toward an improved calibration of the local extragalactic distance scale.

\acknowledgements

     WPG was supported by research grant Fondecyt No. 1971076 which is 
gratefully acknowledged. We are grateful to J.A.R. Caldwell and N. Tanvir for 
making their LMC Cepheid databases available to us, and we appreciate useful 
correspondence with C. Turon who sent us important preprints on the 
Hipparcos results.

\newpage
\figcaption[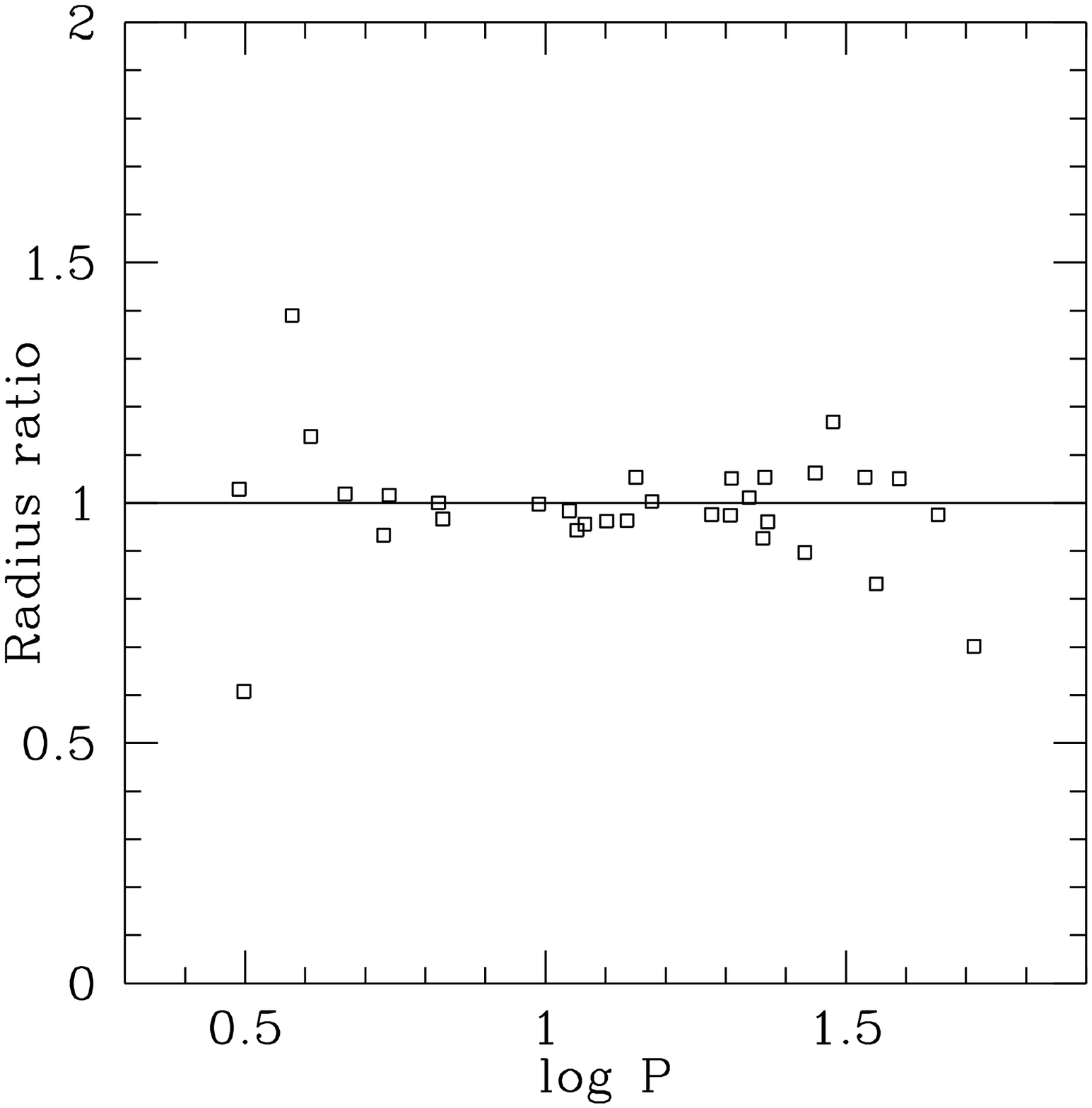]{The ratio of the radii obtained from the $V$, $V-K$ 
   version of our technique to those obtained from the pure infrared $K$, $J-K$
   version, plotted against the pulsation period. The mean ratio is 1.00 with 
   a very low uncertainty, and there is no dependence on period. \label{fig1}}

\figcaption[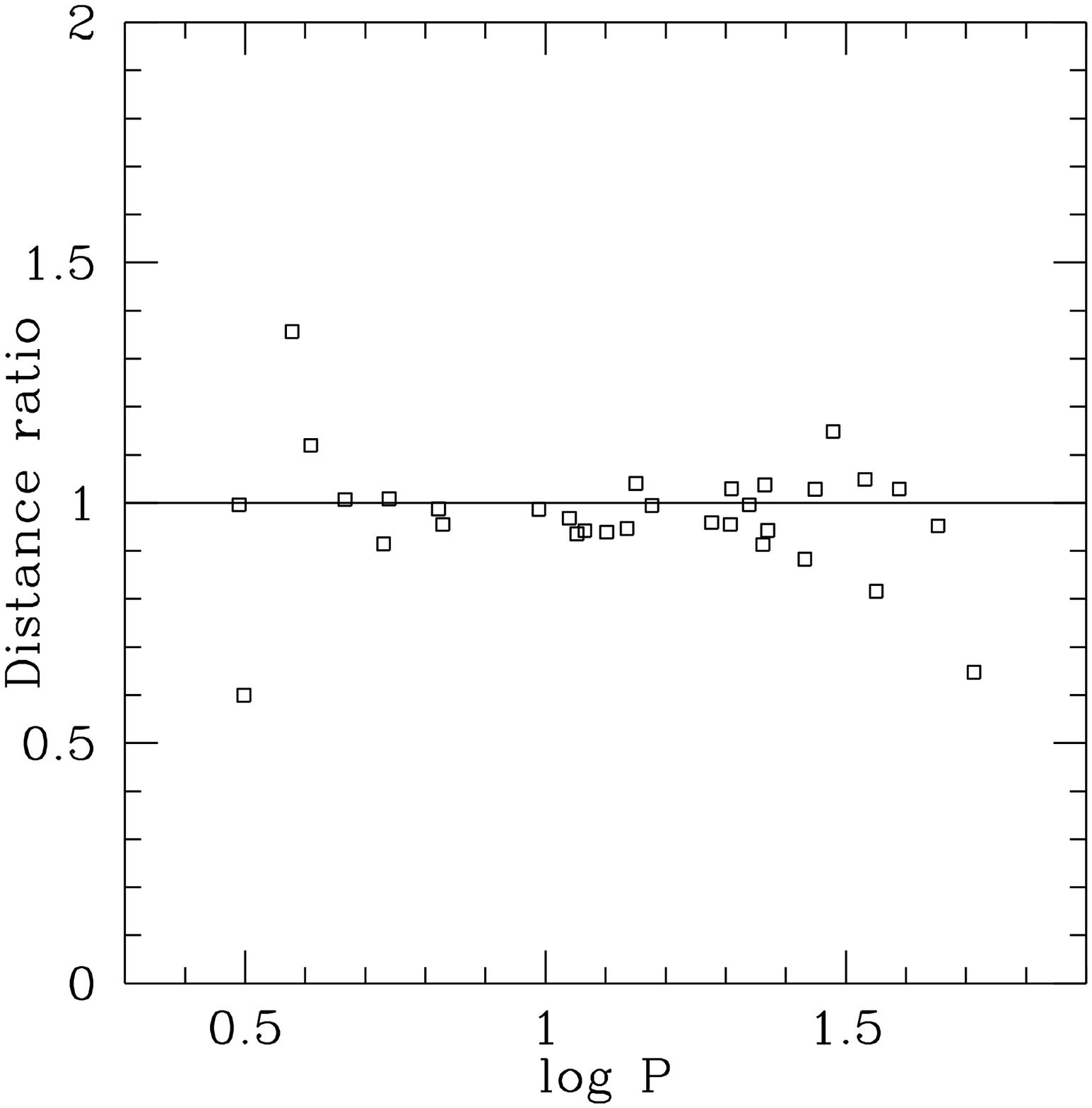]{As Fig.~\ref{fig1}, but for the distances. The mean ratio
   is $0.98 \pm 0.02$, and there is no dependence on period. The slightly 
   increased scatter toward the longer periods is probably due to increased 
   problems with the correct phase alignment between the $V$ and $K$ light 
   curves for the longest-period stars in the sample, which show an enhanced 
   tendency for period variability. \label{fig2}}

\figcaption[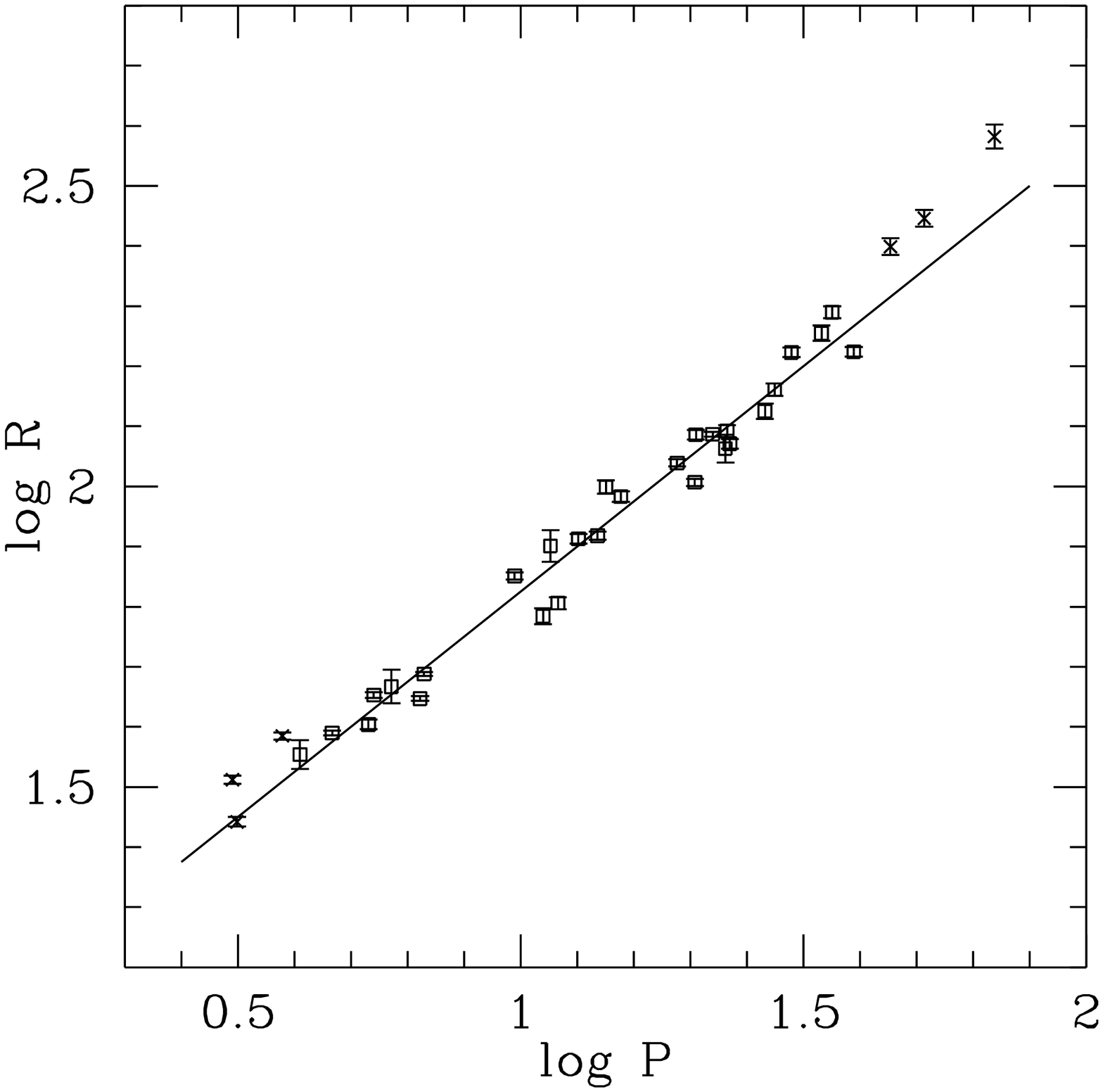]{The period-radius relation defined by 28 Galactic Cepheid
   variables. The plotted line is a least-squares fit to the data. For the
   sake of completeness, excluded stars at short and long periods (see text) 
   are added with a different symbol (cross). \label{fig3}}

\figcaption[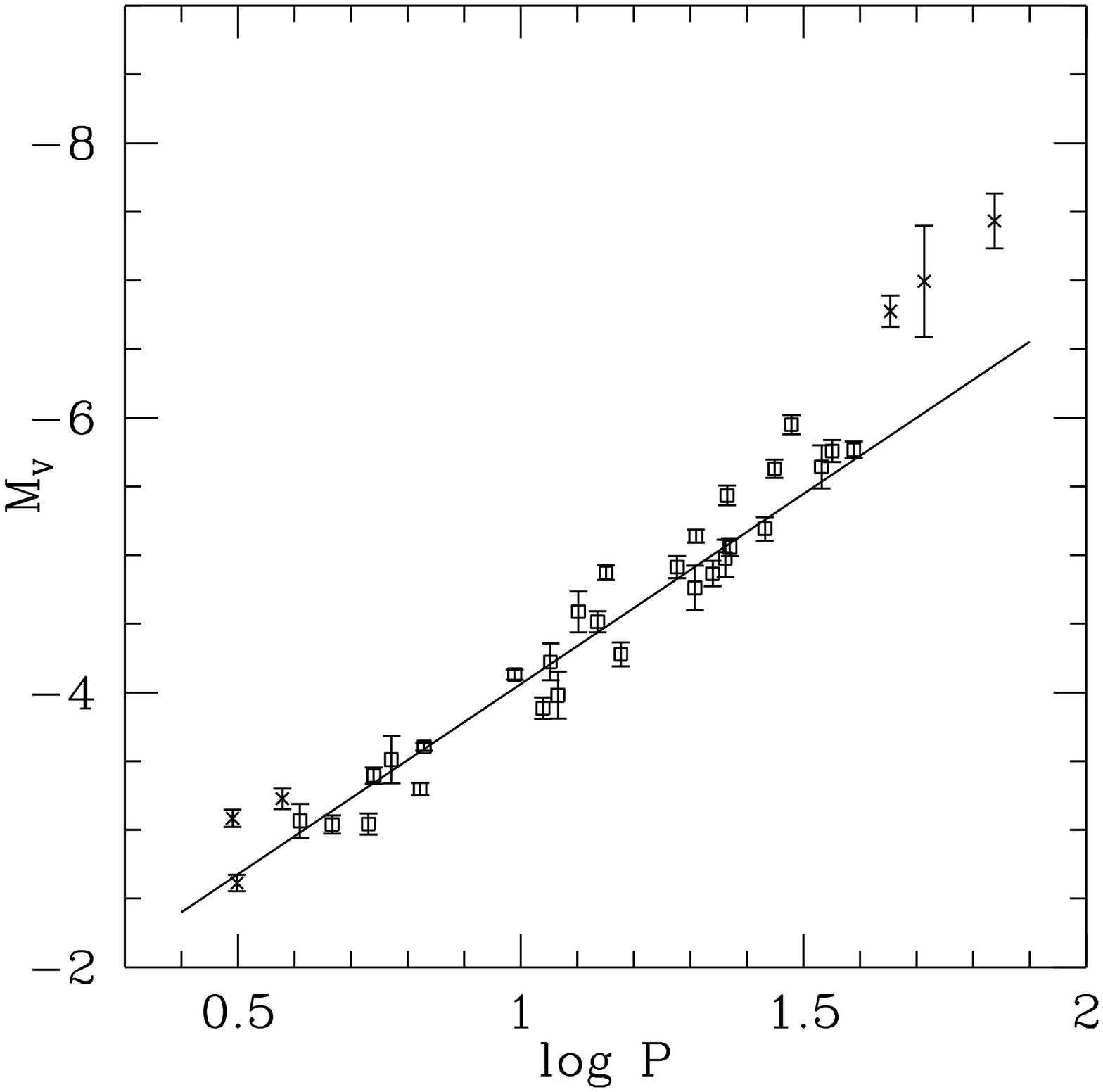]{The $V$ band period-luminosity relation defined by the 
   infrared Barnes-Evans distances of 28 Galactic Cepheids. The error bars 
   correspond to the combined effect of distance and absorption uncertainties  
   on the absolute magnitudes. The plotted line has the slope obtained from a 
   sample of LMC Cepheids (see text and Table~\ref{tab10}). Excluded stars at 
   short and long periods (see text) are added with a different symbol (cross).
   \label{fig4}}

\figcaption[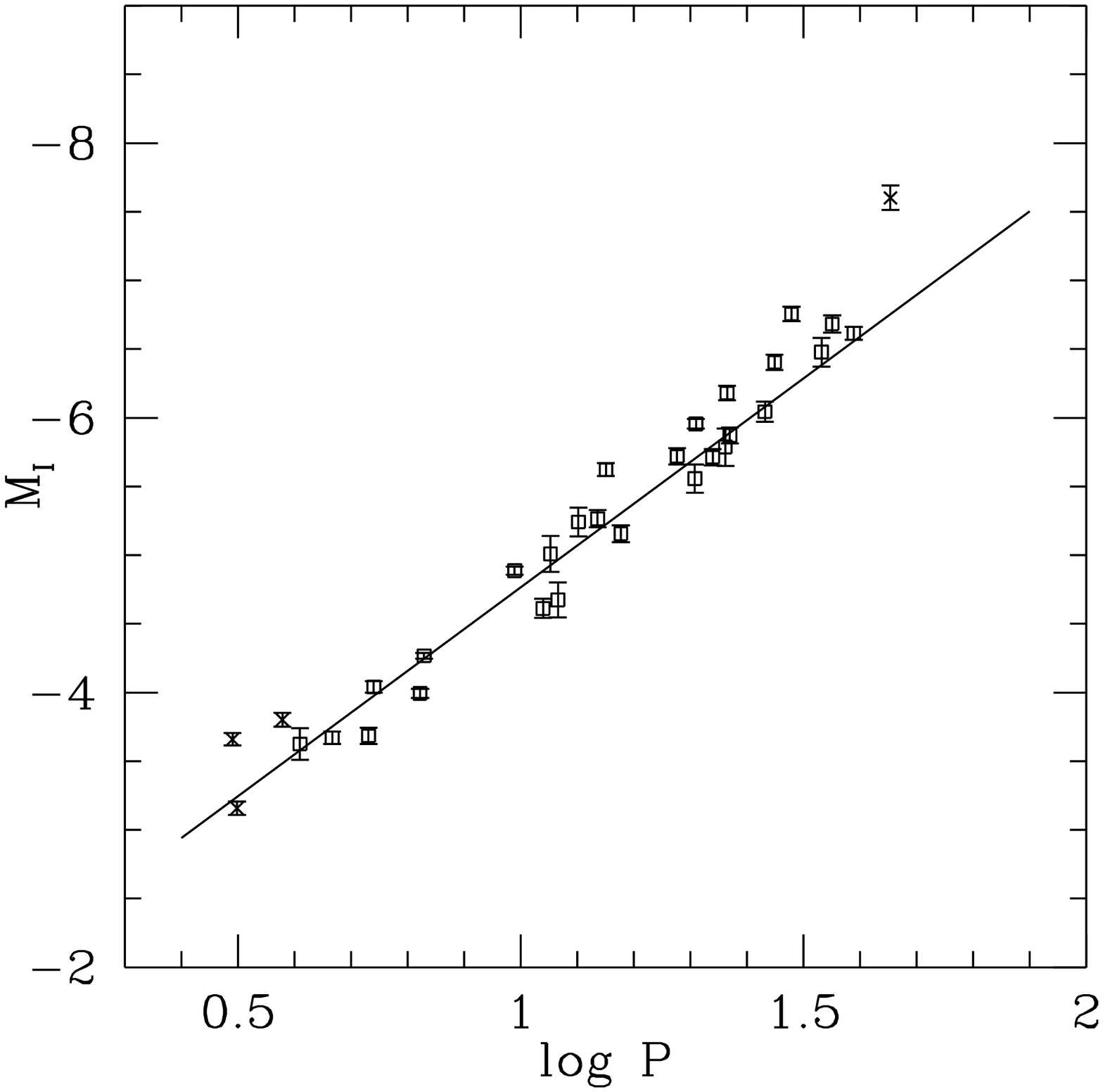]{As Fig.~\ref{fig4}, for the $I$ (Cousins system) 
   passband. \label{fig5}}

\figcaption[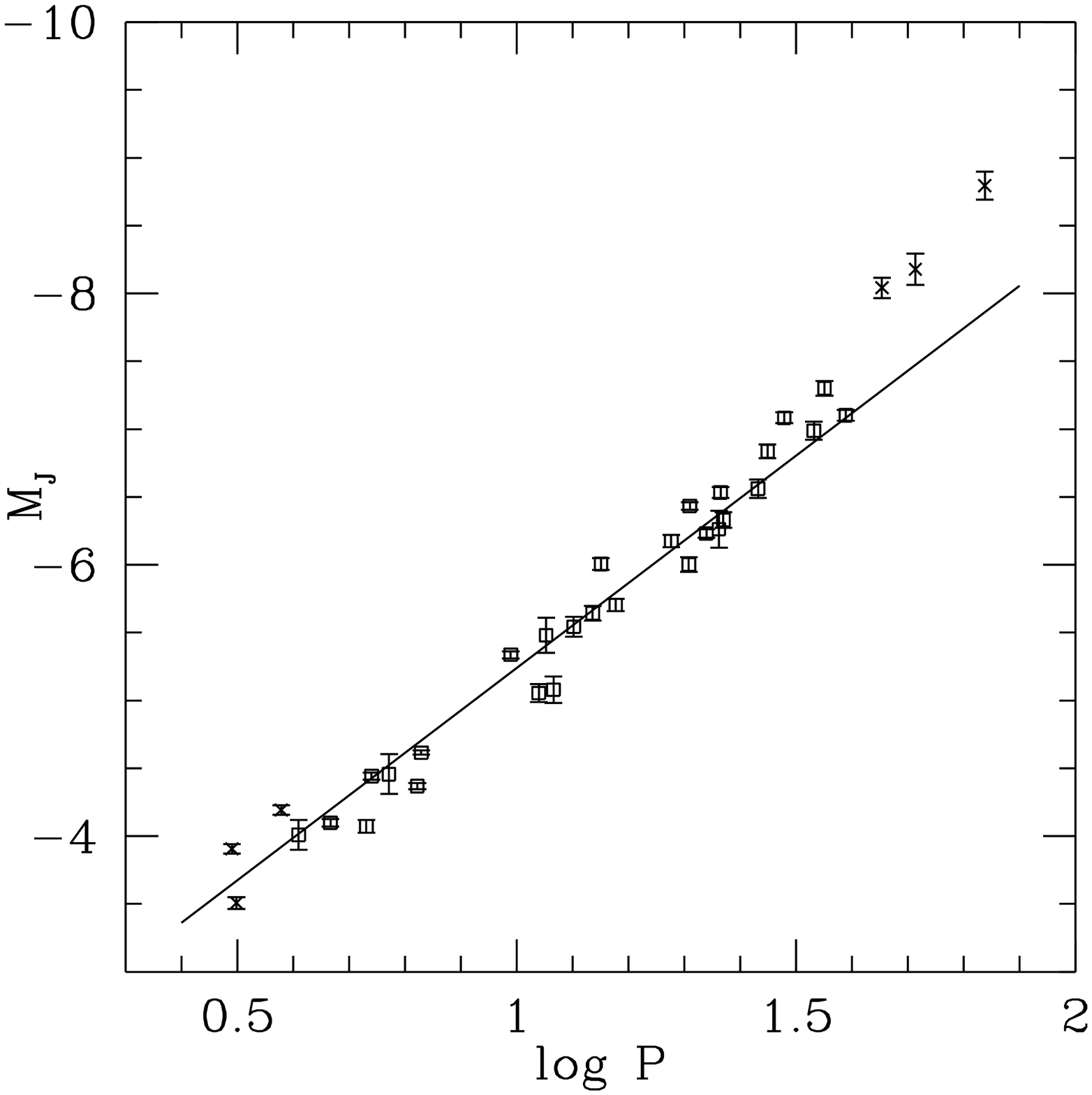]{As Fig.~\ref{fig4}, for the $J$ (Carter system) passband.
   \label{fig6}}

\figcaption[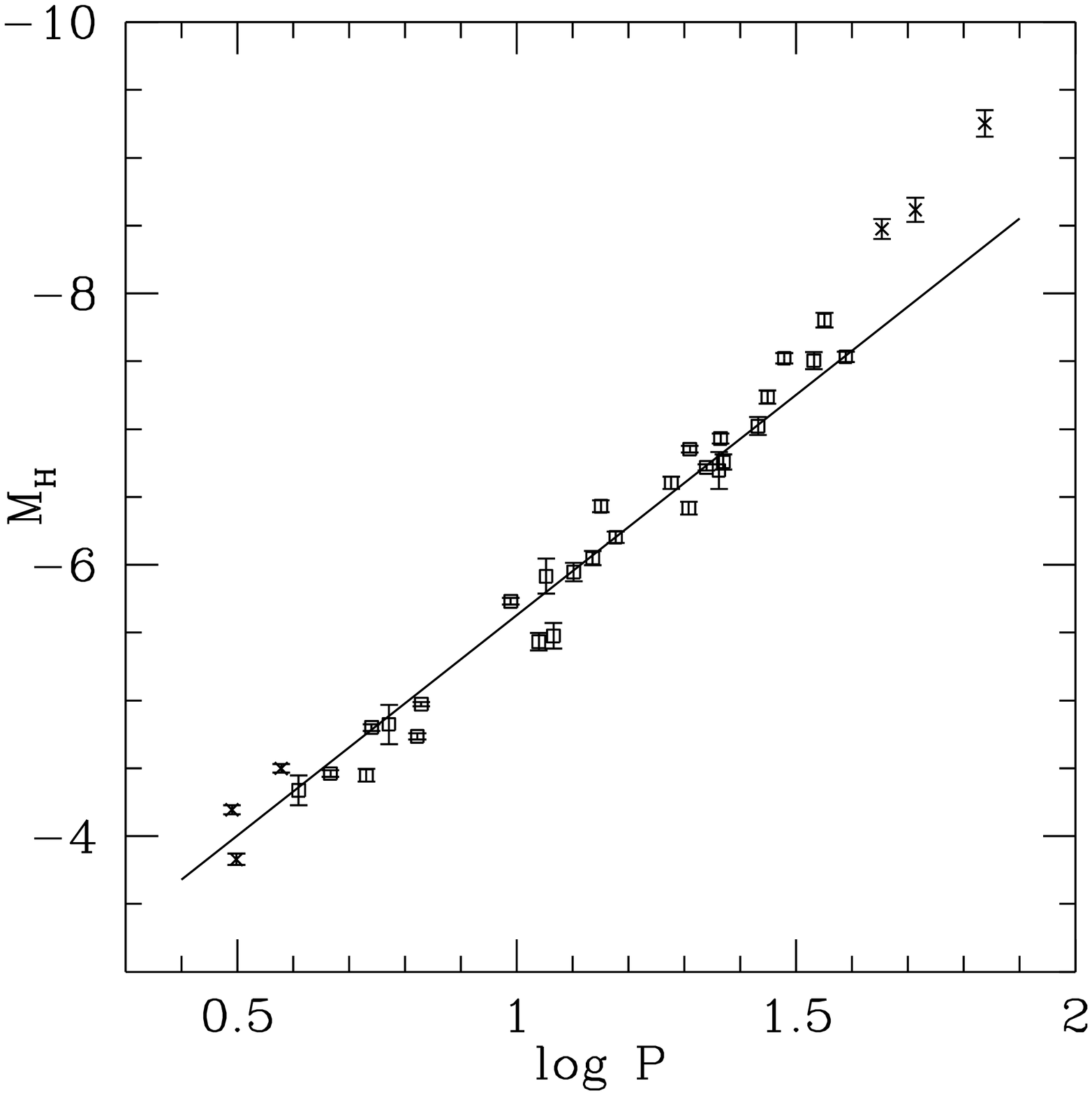]{As Fig.~\ref{fig4}, for the $H$ (Carter system) passband.
   \label{fig7}}

\figcaption[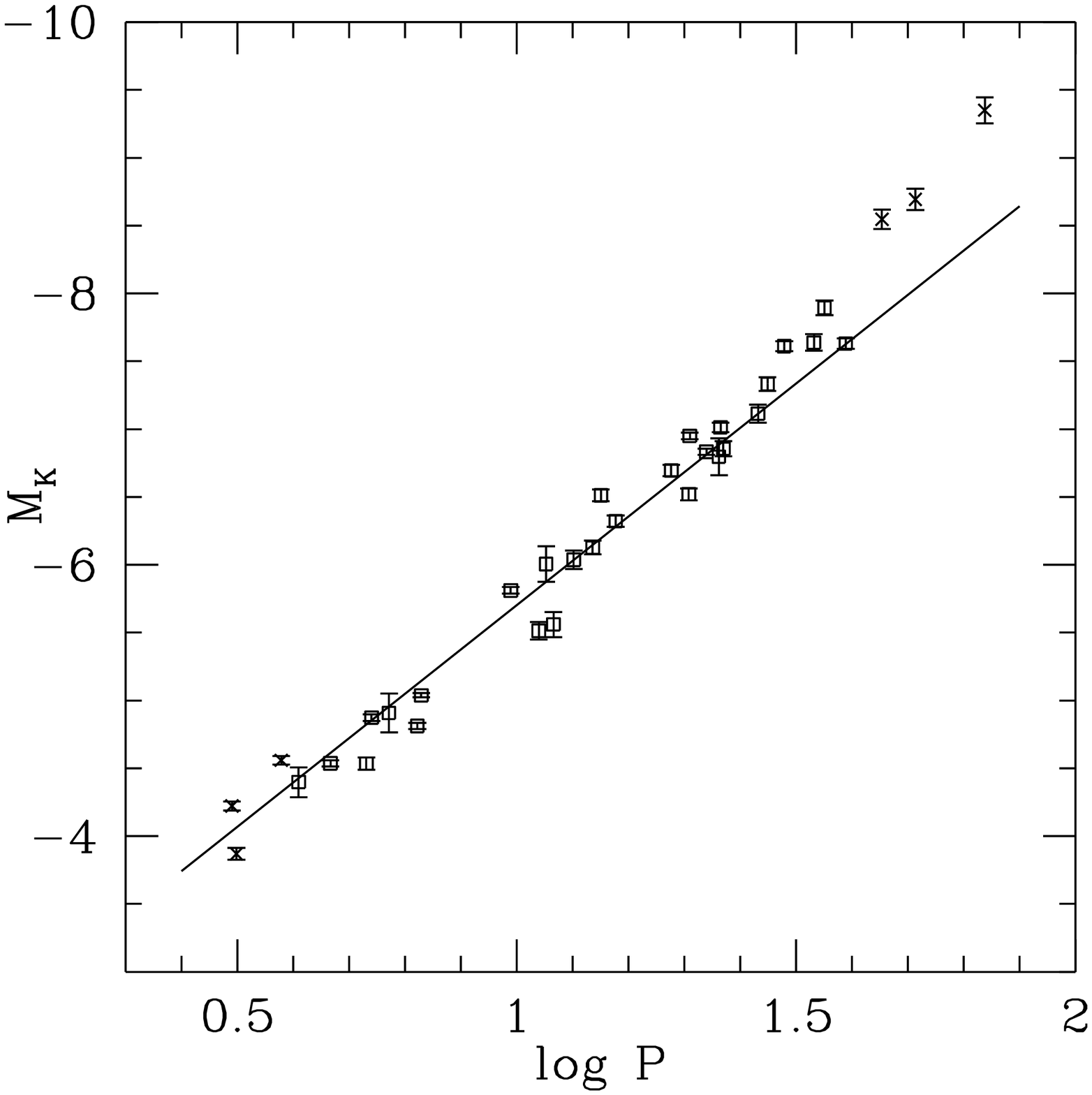]{As Fig.~\ref{fig4}, for the $K$ (Carter system) passband.
   \label{fig8}}

\figcaption[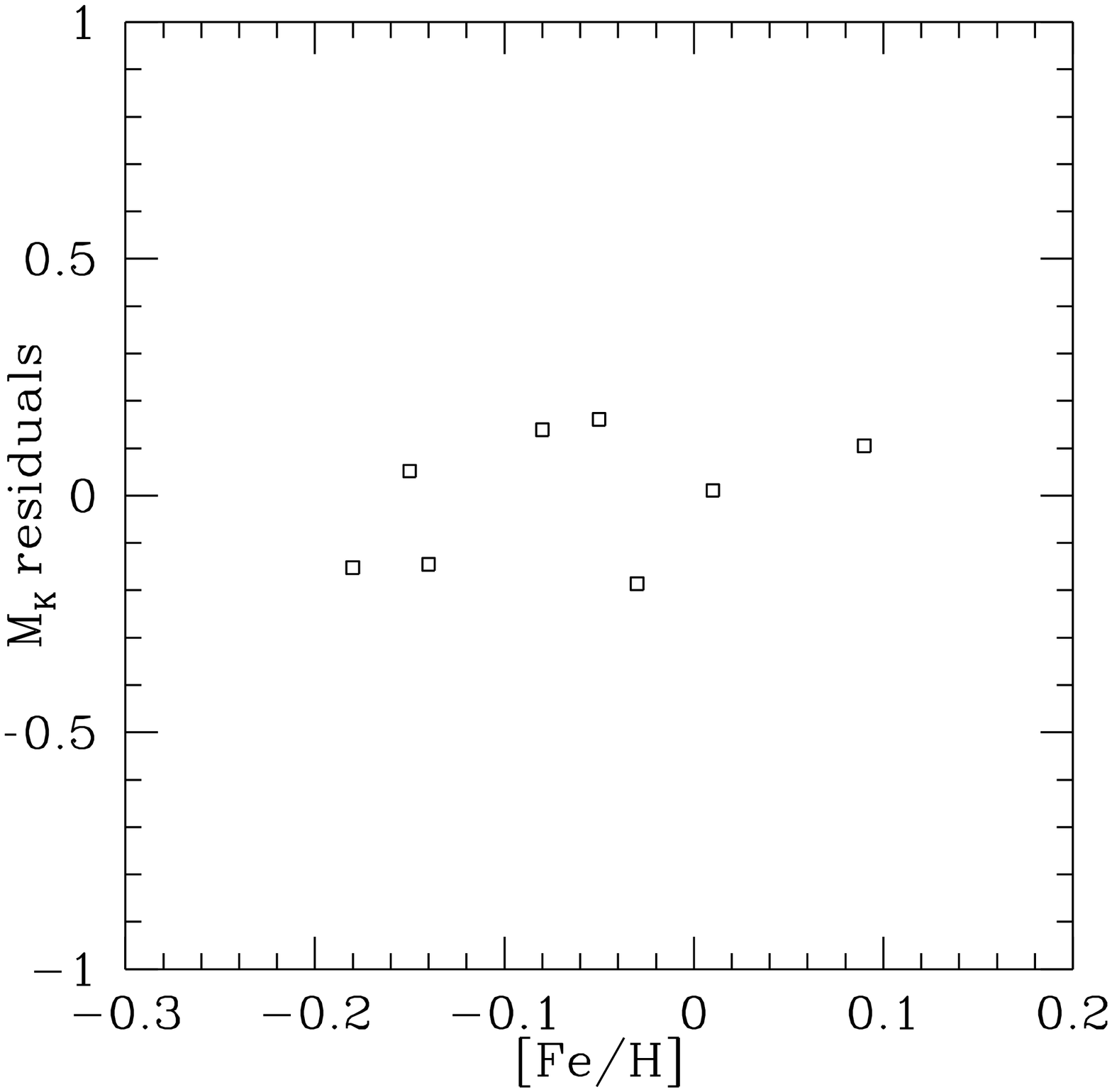]{The residuals of the $K$ band absolute magnitudes from a 
   least-squares fit to the Galactic Cepheid $K$ band PL relation, plotted 
   against the metallicities as determined by Fry \& Carney, for 8 Cepheids 
   common to both samples. There is no correlation. \label{fig9}}


\begin{references}

\reference{bel89} Bell, R. A., \& Gustafsson, B. 1989, \mnras, 236, 653
\reference{ber97a} Berdnikov, L. N. 1997, 
   in ``http://www.physics.mcmaster.ca/Cepheid/''
\reference{ber97b} Berdnikov, L. N. 1997, in Proc. of A Half Century of Stellar
   Pulsation Interpretations: A Tribute to Arthur N. Cox, (Los Alamos: ASP 
   Conf. Ser.), in press
\reference{ber95} Berdnikov, L. N., \& Turner, D. G. 1995, Astronomy Letters, 
   Vol. 21, No. 6, 717
\reference{ber94} Bersier, D., Burki, G., Mayor, M., \& Duquennoy, A. 1994, 
   \aaps, 108, 25
\reference{cac81} Caccin, B., Onnembo, A., Russo, G., \& Sollazzo, C. 1981, 
   \aap, 97, 104
\reference{cal87} Caldwell, J. A. R., \& Coulson, I. M. 1987, \aj, 93, 1090
\reference{cal91} Caldwell, J. A. R., \& Laney, C. D. 1991, in IAU Symp. 148, 
   The Magellanic Clouds, ed. R. Haynes \& D. Milne, (Dordrecht: Kluwer), 249
\reference{cou85a} Coulson, I. M., \& Caldwell, J. A. R. 1985, SAAO Circ., 9, 5
\reference{cou85b} Coulson, I. M., Caldwell, J. A. R., \& Gieren, W. P. 1985, 
   \apjs, 57, 595
\reference{fea97} Feast, M. W., \& Catchpole, R. M. 1997, \mnras, 286, L1
\reference{fer97a} Fernie, J. D., Beattie, B., Evans, N. R., \& Seager, S. 
   1997, ``http://ddo.astro.utoronto.ca/cepheids.html''
\reference{fer97b} Fernley, J., Barnes, T. G., Skillen, I., Hawley, S. L., 
   Hanley, C. J., Evans, D. W., Solano, E., \& Garrido, R. 1997, in Proc. of A 
   Half Century of Stellar Pulsation Interpretations: A Tribute to Arthur N. 
   Cox, (Los Alamos: ASP Conf. Ser.), in press
\reference{fou97} Fouqu\'e, P., \& Gieren, W. P. 1997, \aap, 320, 799 (Paper~I)
\reference{fry97} Fry, A. M., \& Carney, B. W. 1997, \aj, 113, 1073
\reference{gie81a} Gieren, W. P. 1981a, \apjs, 46, 287
\reference{gie81b} Gieren, W. P. 1981b, \apjs, 47, 315
\reference{gie85} Gieren, W. P. 1985, \apj, 295, 507
\reference{gie89a} Gieren, W. P. 1989, \aap, 225, 381
\reference{gie89b} Gieren, W. P., Barnes, T. G., \& Moffett, T. J. 1989, 
   \apj, 342, 467
\reference{gie93a} Gieren, W. P., Barnes, T. G., \& Moffett, T. J. 1993, 
   \apj, 418, 135
\reference{gie93b} Gieren, W. P., \& Fouqu\'e, P. 1993, \aj, 106, 734
\reference{gie97} Gieren, W. P., Fouqu\'e, P., \& G\'omez, M. 1997, 
   \apj, in press (Paper~II)
\reference{gir86} Giridhar, S. 1986, J. Astrophys. Astron., 7, 83
\reference{gor92} Gorynya, N. A., Irsmambetowa, T. R., Rastorgouev, A. S., 
   \& Samus, N. N. 1992, Astronomy Letters, Vol. 18, 5
\reference{gor96} Gorynya, N. A., Samus, N. N., Rastorgouev, A. S., 
   \& Sachkov, M.E. 1996, Astronomy Letters, Vol. 22, No. 2, 175
\reference{gou97} Gould, A., \& Uza, O. 1997, \apj, in press (astro-ph/9705051)
\reference{kro96} Krockenberger, M., Sasselov, D. D., \& Noyes, R. W. 1996, 
   \apj, in press (astro-ph/9611123)
\reference{lan92} Laney, C. D., \& Stobie, R. S. 1992, \aaps, 93, 93
\reference{lan93} Laney, C. D., \& Stobie, R. S. 1993, \mnras, 263, 921
\reference{lan94} Laney, C. D., \& Stobie, R. S. 1994, \mnras, 266, 441
\reference{lan95} Laney, C. D., \& Stobie, R. S. 1995, \mnras, 274, 337
\reference{llo80} Lloyd Evans, T. 1980, SAAO Circ., Vol. 1, No. 5, 257
\reference{mad91} Madore, B. F., \& Freedman, W. L. 1991, \pasp, 103, 933
\reference{mad97} Madore, B. F., \& Freedman, W. L. 1997, \apj, in press 
   (astro-ph/9707091)
\reference{mer87} Mermilliod, J.-C., Mayor, M., \& Burki, G. 1987, 
   \aaps, 70, 389
\reference{mer97} Mermilliod, J.-C., Turon, C., Robichon, N., Arenou, F., 
   \& Lebreton, Y., in Proceedings of the Hipparcos, Venice 1997 Symposium, 
   ESA-SP 402, ed. , (: European Space Agency), in press
\reference{met91} Metzger, M. R., Caldwell, J. A. R., McCarthy, J. K., 
   \& Schechter, P. L. 1991, \apjs, 76, 803
\reference{met92} Metzger, M. R., Caldwell, J. A. R., \& Schechter, P. L. 1992,
   \aj, 103, 529
\reference{mof84} Moffett, T. J., \& Barnes, T. G. 1984, \apjs, 55, 389
\reference{mof97} Moffett, T. J., Gieren, W. P., Barnes, T. G., \& G\'omez, M. 
   1997, \apjs, submitted
\reference{mus97} Musella, I., Piotto, G., \& Capaccioli, M. 1997, 
   \aap, in press (astro-ph/9706285)
\reference{pel76} Pel, J. W. 1976, \aaps, 24, 413
\reference{pon94} Pont, F., Burki, G., \& Mayor, M. 1994, \aaps, 105, 165
\reference{qui94} Quintana, H., Fouqu\'e, P., \& Way, M. J. 1994, 
   \aap, 283, 722
\reference{rip97} Ripepi, V., Barone, F., Milano, L., \& Russo, G. 1997, 
   \aap, 318, 797
\reference{sas97} Sasselov, D. D., Beaulieu, J.-P., Renault, C., Grison, P., 
   Ferlet, R., Vidal-Madjar, A., Maurice, E., Pr\'evot, L., Aubourg, E., 
   Bareyre, P., Brehin, S., Coutures, C., Palanque, N., de Kat, J., Gros, M., 
   Laurent, B., Lachi\`eze-Rey, M., Lesquoy, E., Magneville, C., Milsztajn, A.,
   Moscoso, L., Queinnec, F., Rich, J., Spiro, M., Vigroux, L., Zylberajch, S.,
   Ansari, R., Cavalier, F., Moniez, M., Gry, C., Guibert, J., Moreau, O., 
   Tajhmady, F.,  1997, \aap, 324, 471
\reference{seb95} Sebo, K. M., \& Wood, P. R. 1995, \apj, 449, 164
\reference{sto88} Stothers, R. B. 1988, \apj, 329, 712
\reference{tan97} Tanvir, N. R. 1997, in STScI Symp. Series, Vol. 10, The 
   Extragalactic Distance Scale, ed. M. Livio, M. Donahue, \& N. Panagia, 
   (Cambridge: Cambridge University Press), 91
\reference{vle97} van Leeuwen, F., \& Hansen Ruiz, C. S. 1997, in Proceedings 
   of the Hipparcos, Venice 1997 Symposium, ESA-SP 402, ed. , 
   (: European Space Agency), in press
\reference{wel94} Welch, D. L. 1994, \aj, 108, 1421
\reference{wel97} Welch, D. L. 1997, 
   ``http://www.physics.mcmaster.ca/Cepheid/''
\reference{wel87} Welch, D. L., McLaren, R. A., Madore, B. F., 
   \& McAlary, C. W. 1987, \apj, 321, 162
\end{references}
\end{document}